\documentclass[
]{ceurart}

\usepackage{listings}
\usepackage[all]{nowidow}
\begin{document}

\copyrightyear{2023}
\copyrightclause{Copyright for this paper by its authors.
  Use permitted under Creative Commons License Attribution 4.0
  International (CC BY 4.0).}

\conference{CLEF 2023: Conference and Labs of the Evaluation Forum, September 18–21, 2023, Thessaloniki, Greece}

\title{Evaluating Temporal Persistence Using Replicability Measures}

\title[mode=sub]{Notebook for the LongEval Lab at CLEF 2023}



\address[1]{TH Köln (University of Applied Sciences),
  Claudiusstr. 1, Cologne, 50678, Germany}

\author[1]{Jüri Keller}[%
orcid=0000-0002-9392-8646,
email=jueri.keller@smail.th-koeln.de
]
\cormark[1]

\author[1]{Timo Breuer}[%
orcid=0000-0002-1765-2449,
email=timo.breuer@th-koeln.de
]

\author[1]{Philipp Schaer}[%
orcid=0000-0002-8817-4632,
email=philipp.schaer@th-koeln.de
]

\cortext[1]{Corresponding author.}

\begin{abstract}
    In real-world Information Retrieval (IR) experiments, the Evaluation Environment (EE) is exposed to constant change. Documents are added, removed, or updated, and the information need and the search behavior of users is evolving. Simultaneously, IR systems are expected to retain a consistent quality.
    The LongEval Lab seeks to investigate the longitudinal persistence of IR systems, and in this work, we describe our participation. We submitted runs of five advanced retrieval systems, namely a Reciprocal Rank Fusion (RRF) approach, ColBERT, monoT5, Doc2Query, and E5, to both sub-tasks.
    Further, we cast the longitudinal evaluation as a replicability study to better understand the temporal change observed.
    As a result, we quantify the persistence of the submitted runs and see great potential in this evaluation method.
\end{abstract}


\maketitle

\section{Introduction}
This paper describes our contribution to the CLEF 2023 LongEval Lab~\cite{longevaloverview2023}.\footnote{\url{https://clef-longeval.github.io}} The lab seeks to investigate the temporal persistence of retrieval systems. It, therefore, provides a first-of-its-kind web retrieval collection with three sub-collections from different points in time~\cite{deveaudLongEvalRetrievalFrenchEnglishDynamic2023}. We participated in the retrieval task by providing runs of five systems to both sub-task. 

A retrieval system's Evaluation Environment (EE) is under constant change. Not only but especially web retrieval systems are exposed to this due to the dynamic nature of the web. Documents, i.e., websites, get created, updated, or created~\cite{bar-ilanCriteriaEvaluatingInformation2002, dumaisTemporalDynamicsInformation2010}. But besides the evolving collection, all other aspects of an EE underlay change as well, from the information need and search behavior of the users~\cite{adarWebChangesEverything2009} all the way to the evolving language itself~\cite{jatowtLargeScaleAnalysis2012}.
These changes raise questions about the persistence and generalizability of IR system effectiveness evaluations.

By requiring a temporarily reliable system to perform consistently over time, evaluating this can be understood as a replicability task. Oriented at the ACM definition of replicability\footnote{\url{https://www.acm.org/publications/policies/artifact-review-and-badging-current}}, the goal is to achieve the same measurements in a different experimental setup, in this case, at a proceeded point in time. 

To investigate temporal persistence, we submitted runs of five advanced retrieval systems to both sub-tasks of the LongEval Lab. The systems are not specifically adapted to changes in the LongEval dataset to validate the temporal reliability of system-oriented IR evaluations following the Cranfield paradigm. Further, as a proof of concept, we use the replicability measures Delta Relative Improvement ($\Delta$~RI) and the Effect Ratio (ER)~\cite{breuerHowMeasureReproducibility2020a} to investigate the temporal persistence.
In short, the contributions of this work are:

\begin{itemize}
    \item Descriptions of \textbf{five state-of-the-art systems} submitted to both retrieval sub-tasks,
    \item an \textbf{extensive evaluation} of retrieval effectiveness,
    \item an \textbf{adaptation of replicability measures} to evaluate temporal persistence,
    \item an \textbf{open-source release} of the experimental setup. 
\end{itemize}

The remainder of this paper is structured as follows. Section~\ref{sec:LongEval_Dataset} contains an analysis of the LongEval dataset. The five retrieval systems are described in Section~\ref{sec:approaches-and-implementations}. Further, Section~\ref{sec:evaluation} provides the results on the train slice and a preliminary evaluation of the results. In Section~\ref{sec:replicability}, we describe the replicability efforts. This paper concludes with a short discussion and some future work in Section~\ref{sec:conclusion}. The code is publicly available on GitHub.\footnote{\url{https://github.com/irgroup/CLEF2023-LongEval-IRC}}

\section{LongEval Dataset}\label{sec:LongEval_Dataset}
To our knowledge, the LongEval dataset~\cite{deveaudLongEvalRetrievalFrenchEnglishDynamic2023} is the first dataset specifically designed to investigate temporal changes in IR. On a high level, the collection consists of three sub-collections from different points in time. Each collection contains topics and qrels. The documents as well as the topics and qrels originate from the French, privacy-focused search engine Qwant.\footnote{\url{https://www.qwant.com/}} For this work, we entirely rely on the English automatic translations of the dataset. 
The documents contain the cleaned content of websites. They are filtered for adult and spam content, but no further processing was done, sometimes leaving unconnected phrases, keywords, or code artifacts in the documents.

The topics are selected according to \emph{``popularity, stability, generality, and diversity''}~\cite{deveaudLongEvalRetrievalFrenchEnglishDynamic2023}. For these topics, queries are selected from the Qwant search engine logs if they contain the topic as a sub-string.
The qrels for the shared task are simulated based on the Cascade Click Model~\cite{DBLP:conf/www/ChapelleZ09, DBLP:conf/wsdm/CraswellZTR08}. Documents are assessed as not relevant, relevant, and highly relevant. Further, human-assessed gold labels are announced for September (2023). More details can be found in the original publication~\cite{deveaudLongEvalRetrievalFrenchEnglishDynamic2023}.

The sub-collections are sequential snapshots of an evolving search environment for temporal comparison. The topics are constructed once, but the queries are partially changing across sub-collections. The documents, i.e., the websites identified by the URL, are also mainly static across sub-collections but the content of the documents changes. 

\begin{table}
  \caption{LongEval subcollection statistics. The length of documents and queries are measured in tokens, split by white spaces. The query WT q062213307 and ST q072211861 is excluded as outlier since it only contains the token \textit{leg} 108 and 110 times.}
  \label{tab:LongEval-dataset}
  \begin{tabular}{lrrrr}
    \toprule
                            & WT   & ST   & LT       & Intersection \\ \midrule
        Timeframe           & June 2022     & July 2022     & September 2022    &                       \\ \midrule
        Number documents    & 1,570,734     & 1,593,376     & 1,081,334         & 1,011,613             \\
        Mean document length& 794.11        & 793.96        & 807.28            &                       \\
        Min document length & 0             & 0             & 1                 &                       \\
        Max document length & 7065          & 12210         & 7255              &                       \\ \midrule
        Number queries      & 753           & 860           & 910               & 124                   \\
        Mean query length   & 2.73          & 2.71          & 2.52              &                       \\
        Min query length    & 1             & 1             & 1                 &                       \\
        Max query length    & 6             & 11            & 9                 &                       \\
    \bottomrule
  \end{tabular}
\end{table}

\begin{figure}
  \centering
  \includegraphics[width=\linewidth]{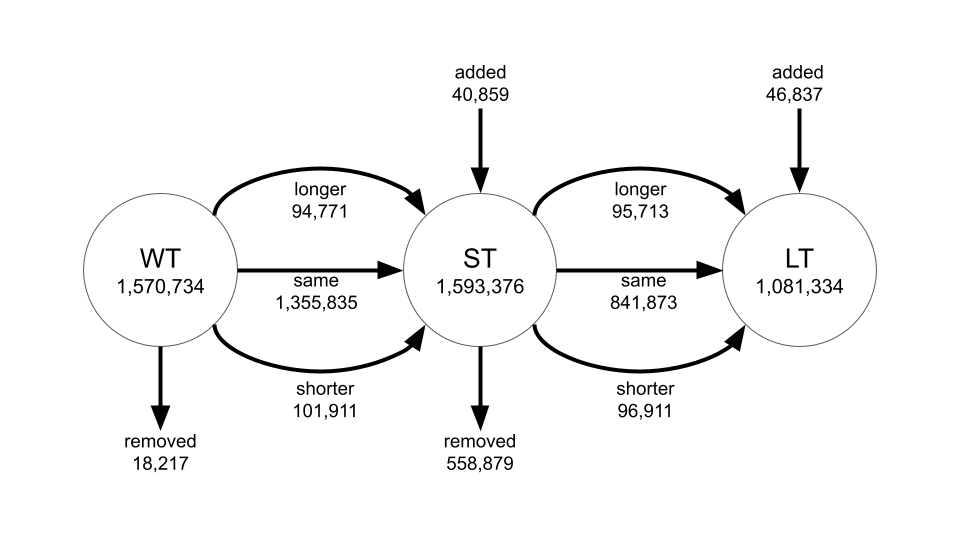}
  \caption{The evolution of the LongEval dataset documents across the three sub-collections. Transitioning from one sub-collection to the next, documents are added, removed, or updated. All documents were harmonized by their URLs.}\label{fig:longeval-doc-evolution}
\end{figure}

The collections are organized into a WT, ST, and LT sub-collection. The WT (within time) sub-collection was created in June 2022. The ST (short-term) sub-collection was created in July~2022, immediately after the WT collection. The third sub-collection, LT (long term), contains more distant data as it was created with a two-month gap from ST in September 2022. Table~\ref{tab:LongEval-dataset} gives an overview of the sub-collections.

The LongEval dataset contains over 1.5 million documents. Not every document is present in every sub-collection, but most documents are. The core document collection contains 1,011,613 documents, as identified by matching their URLs. Versions of these documents are present in every sub-collection but do not necessarily contain exactly the same content. The documents evolve over time, meaning that the content of one website might change over time. To capture this change on a general level, Figure~\ref{fig:longeval-doc-evolution} shows how many documents increase or decrease in character length and how many documents are added, deleted, or stay the same in length. We note that between ST and LT considerably more documents are removed from the collections than between WT and ST.

Like the documents, the queries change over time as well. However, relatively fewer core queries that appear in all sub-collections exist. In total, only 124 unique query strings appear in all collections. However, the overlap of query IDs is larger due to duplicate queries that are probably caused by automatic translations.

\begin{figure}
  \centering
  \includegraphics[width=\linewidth]{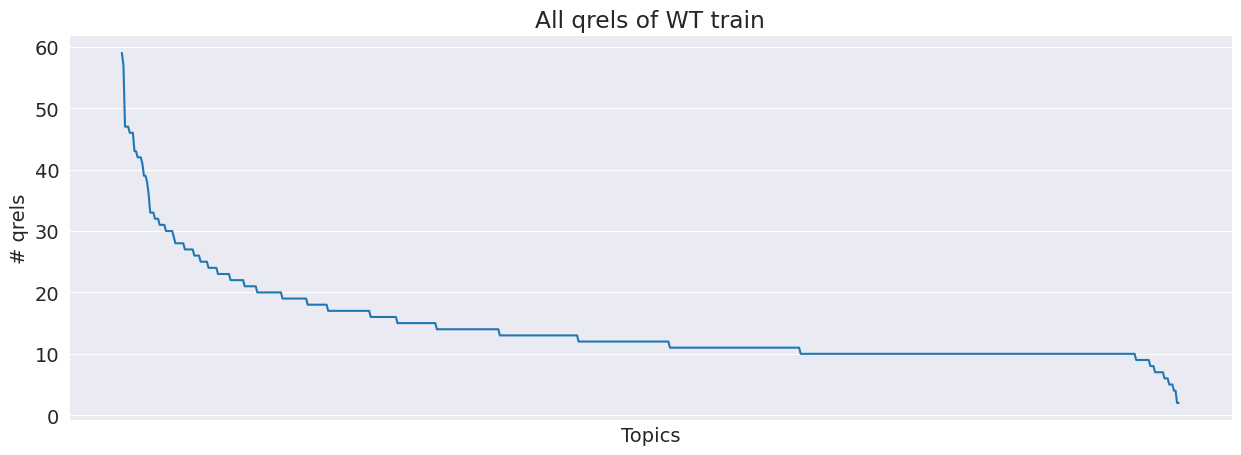}
  \caption{Distribution of qrels per query for the 672 WT train sub-collection queries.}\label{fig:longeval-qrels}
\end{figure}

The relevance judgments (qrels) classify  the documents' relevance on a three-graded scale, including \textit{not relevant}, \textit{relevant}, and \textit{highly relevant} labels. In general, the dataset has few assessed documents per topic. While the mean number of qrels is 14 per topic, the absolute number fluctuates between 2 and 59. Figure~\ref{fig:longeval-qrels} shows the distribution of all qrels per query. 
Most of the documents are marked as not relevant, and the distribution of relevant and highly relevant qrels is skewed as well. Especially the highly relevant qrels are rare, with a maximum of only four and a mean of only one highly relevant document per topic. In the evaluations, these single documents heavily influence the final outcome as their position in the ranking  especially impacts the score of rank-based measures like nDCG. While relevant qrels are generally rare, 16 queries do not have a single relevant document. 

\section{Approaches and Implementations}\label{sec:approaches-and-implementations}
We compared different ranking functions and multi-stage retrieval systems on the WT train slice of the LongEval dataset. The systems were chosen as they represent state-of-the-art, off-the-shelf methods that are used in many recent IR experiments. Therefore, it is especially interesting how these systems behave over time without being specifically adapted to a changing environment.

\subsection{Statistical Ranking Functions}
Different ranking functions were used as baselines in their default configurations. Special attention was given to the BM25~\cite{robertsonOkapiTREC31994} ranking function as it is a robust, efficient, and often hard-to-beat baseline. We use this run to compare advanced systems to it. Since we use the PyTerrier~\cite{macdonaldDeclarativeExperimentationInformation2020} framework for experiments, the default parameters $k_1 = 1.2$ and $b = 0.75$ were kept. Further we included PL2~\cite{amatiProbabilityModelsInformation2003}, TF-IDF and DFR $\chi^2$~\cite{DBLP:conf/ecir/Amati06}.

To further improve these ranking functions, two query expansion methods are employed. Namely, RM3~\cite{jaleelUMassTREC20042004} and Bo1~\cite{amatiProbabilityModelsInformation2003} are used to extend the queries through pseudo-relevance feedback. 
The default PyTerrier parameters are also kept here; three feedback documents were used to gather ten feedback terms.

\subsection{Rank Fusion}
Multiple runs were combined into a single ranking to profit from the diversity of multiple ranking functions. First, BM25, DFR $\chi^2$ and PL2 are fused through Reciprocal Rank Fusion (RRF)~\cite{cormackReciprocalRankFusion2009} with the \texttt{ranx} Python library~\cite{bassaniRanxFusePython2022}. Further runs are created by using the pseudo-relevance-feedback methods on top of BM25. The default parameters $min_k = 10$, $max_k = 100$ and $step = 10$ were used for the RRF.

\subsection{ColBERT}
ColBERT~\cite{khattabColBERTEfficientEffective2020} applies the BERT~\cite{devlinBERTPretrainingDeep2019} Language Model (LM) to overcome the lexical gap~\cite{furnas_vocabulary_1987} by creating semantic representations of queries and documents as embeddings. In contrast to traditional BERT-based approaches like cross-encoders, the interaction mechanism used to calculate the similarity between a document and a query is detached from the embedding creation process. However, in contrast to bi-encoder systems, nuanced similarities can be calculated. To do so, semantic representations for a query or a document are calculated as a set of token embeddings. The relevance score between a query and a document is then calculated as the sum of the max of the cosine similarity or the L2 distance between all embeddings for the query and the document.

By separating the scoring from the embedding process, the efficiency at run time can be greatly improved as all document embeddings can be calculated beforehand offline. ColBERT can also be used in a later retrieval stage as a reranker.
The PyTerrier version of ColBERT~\footnote{\url{https://github.com/terrierteam/pyterrier_colbert}} was used in a zero-shot fashion. Besides using ColBERT as a first-stage retriever, where the whole corpus is converted to embeddings, ColBERT was also used to rerank the top 1000 BM25 results. 

\subsection{monoT5}
The potential of sequence-to-sequence models can be fostered for the ranking task by providing a query and a document as input and asking the model to decide if the document is relevant for this query by generating "true" or "false." The softmax of the generated token probability is then used as confidence for the predicted class to compute the final relevance of the document~\cite{nogueiraDocumentRankingPretrained2020}. The T5~\cite{raffelExploringLimitsTransfer2020} model was fine-tuned in this fashion on the MS Marco passage retrieval dataset~\cite{DBLP:conf/nips/NguyenRSGTMD16} as monoT5 by Pradeep et al.~\cite{pradeepExpandoMonoDuoDesignPattern2021}. This model is then used in a second stage to rerank BM25 rankings and achieves great results, even as a pre-trained model on other datasets and domains~\cite{pradeepExpandoMonoDuoDesignPattern2021}. 

The T5 model supports 512 sub-word tokens, and the LongEval dataset consists of documents with an average length of around 800 tokens. To avoid arbitrary truncation, the document retrieval task is formulated as a passage retrieval task, and the top 1000 BM25 results are split into (still arbitrary but shorter) passages with an overlap half the size of the passage. By that, the whole document texts are reranked by monoT5. Further, the maximum relevance score of all passages from one document is used as the relevance score of the document for the final ranking.

For comparison and to avoid arbitrary sequences, the full documents are used instead as well. This approach seems reasonable since not too much text is cut off from the average document, and the title and introductions with high-level terms, similar to the query terms, are often located at the beginning of a document and are therefore captured by the model. 

\subsection{Doc2Query}
Instead of applying a language model at the reranking stage, Doc2Query~\cite{cheritonDoc2queryDocTTTTTquery2019} uses the T5 model to generate likely queries that a document could answer. These additional queries are then indexed along the document itself. By that, natural language queries can result in exact matches using traditional ranking functions, and alleged relevant terms are boosted. This results in an advanced index that can be efficiently searched independent of methods. 

The effectiveness is highly dependent on the number of queries that are added to the documents during indexing since this determines how much content is added. For this experiment, we used three and ten queries. While Rodrigo and  Lin~\cite{cheritonDoc2queryDocTTTTTquery2019} used up to 80 queries, a maximum of ten queries were chosen to match the available resources. Three queries are the default of the implementation and were used as a lower bound to test the effect.  

\subsection{E5}
Recently Wang et al.~\cite{wangTextEmbeddingsWeaklysupervised2022} achieved superior performance with the E5 model family. It is the first model that outperforms BM25 in a zero-shot retrieval setting on the BEIR~\cite{thakurBEIRHeterogeneousBenchmark2021} benchmark. The performance is attributed to the large and high-quality dataset, the contrastive pre-training and the advanced fine-tuning process. The new paired dataset CCPairs~\cite{thakurBEIRHeterogeneousBenchmark2021} of query passage pairs was used for training. It contains 1.3 billion query document pairs from Reddit, Wikipedia, SemanticScoolar, CommonCrawl, Stack Exchange, and news websites. 

The models E5\textsubscript{small} and E5\textsubscript{base} are used in a zero-shot fashion to create embeddings for all queries and documents. The documents are truncated at 512 sub-word tokens to fit in the model and not split into passages for efficiency. A Faiss\footnote{\url{https://faiss.ai/}} flat index was created from all embeddings, and L2 was used to score the query document similarity.

\section{Evaluation}\label{sec:evaluation}
\begin{table}[]
  \caption{Results on the train slice of the WT sub-collection. The best results per group are highlighted in \textbf{bold}, and significant differences with Bonferroni correction to the BM25 baseline are denoted by an asterisk~($*$).}
  \label{tab:LongEval-WT-train}
\resizebox{\textwidth}{!}{
\begin{tabular}{lllllll}
\toprule
System                     & MAP                & Bpref   & RR      & P@20    & nDCG    & nDCG@20 \\ \midrule
BM25                       & 0.1452             & 0.3245  & 0.2604  & 0.0654  & 0.2884  & 0.2087  \\ \midrule
PL2                        & 0.1408             & \textbf{0.3352}  & 0.2572  & 0.0650  & 0.2884  & 0.2064  \\
TF-IDF                    & \textbf{0.1467}    & 0.3259  & \textbf{0.2637}  & 0.0660  & \textbf{0.2907}  & \textbf{0.2109}  \\
DFR $\chi^2$                  & 0.1428             & 0.3265  & 0.2629  & \textbf{0.0633}  & 0.2871  & 0.2042  \\ \midrule
BM25+Bo1                   & \textbf{0.1470}    & \textbf{0.3341}  & \textbf{0.2534}  & \textbf{0.0661}  & \textbf{0.2922}  & \textbf{0.2075}  \\
BM25+RM3                   & 0.1426  & 0.3295   & 0.2408  & 0.0658  & 0.2867  & 0.2035  \\ \midrule
RRF(BM25, DFR $\chi^2$, PL2)     & 0.1462  & 0.3380*  & 0.2646  & 0.0656  & 0.2967* & 0.2101  \\
RRF(BM25+Bo1, DFR $\chi^2$, PL2) & \textbf{0.1511}    & 0.3466* & \textbf{0.2686}  & 0.0673  & \textbf{0.3040*}  & \textbf{0.2156}  \\
RRF(BM25+RM3, DFR $\chi^2$, PL2)  & 0.1472             & \textbf{0.3472*} & 0.2589  & \textbf{0.0676}  & 0.3008* & 0.2125  \\ \midrule
BM25+passages+monoT5       & 0.1540             & 0.3369  & 0.2743  & 0.0708* & 0.2969  & 0.2196  \\
BM25+monoT5                & \textbf{0.1809*}   & \textbf{0.3494*} & \textbf{0.3216*} & \textbf{0.0768*} & \textbf{0.3208*} & \textbf{0.249*}  \\ \midrule
d2q(3)>BM25                & 0.1578             & \textbf{0.3411}   & 0.2630           & \textbf{0.0752*}   &  0.2940            & 0.2284* \\
d2q(10)>BM25               & \textbf{0.1638*}   & 0.3382            & \textbf{0.2862*} & 0.0707*            &  \textbf{0.3070*}  & \textbf{0.2287*} \\ \midrule
colBERT                    & 0.1652  & 0.3435   & 0.3045* & 0.0689  & 0.2989  & 0.2290  \\
BM25+colBERT               & \textbf{0.1682*}   & \textbf{0.3447}  & \textbf{0.3046*} & \textbf{0.0692}  & \textbf{0.3082}* & \textbf{0.231*}  \\ \midrule
E5\_small                  & 0.1437  & 0.3265   & 0.2705  & 0.0619  & 0.2762  & 0.2039  \\
E5\_base                   & \textbf{0.1545}    & \textbf{0.3483}  & \textbf{0.2826}  & \textbf{0.0634}  & \textbf{0.2910}  & \textbf{0.2128}  \\ \bottomrule
\end{tabular}
}
\end{table}

In the following, results for the initial experiments on the train slice of the WT sub-collection are reported, and the submitted systems are analyzed. Then, the runs and results on the full dataset are described. 
\subsection{System Selection}
 Table~\ref{tab:LongEval-WT-train} gives an extensive overview of the initial experiments. BM25 appeared to be a strong baseline, outperformed only by some systems and most often not statistically significant on all measures. The best runs of the different types were chosen for submission, also with the goal in mind to provide a diverse set of runs for the planned pooled gold annotation~\cite{deveaudLongEvalRetrievalFrenchEnglishDynamic2023}. 

For the official ranking, we submitted to both sub-tasks the five systems: 
\begin{enumerate}
    \item RRF(BM25+Bo1, DFR $\chi^2$, L2) as \textbf{\texttt{IRC\_RRF(BM25+Bo1-XSqrA\_M-PL2)}}
    \item BM25+colBERT as \textbf{\texttt{IRC\_BM25+colBERT}}
    \item BM25+monoT5 as \textbf{\texttt{IRC\_BM25+monoT5}}
    \item d2q(10)>BM25 as \textbf{\texttt{IRC\_d2q(10)>BM25}}
    \item E5\textsubscript{base} as \textbf{\texttt{IRC\_E5\_base}}
\end{enumerate}

The BM25 baseline achieved an nDCG of 0.2884 on the WT train sub-collection slice. A MAP of 0.1452 is reported, but as initially shown in the data analysis in Section~\ref{sec:LongEval_Dataset}, only a few qrels per query are available; we relied on the bpref~\cite{buckleyRetrievalEvaluationIncomplete2004} measure instead. Here, a score of 0.3245 is achieved. Notably, compared to BM25, TF-IDF outperforms BM25 slightly but is not statistically significant. Regarding the runs with additional pseudo-relevance feedback, no significant improvements are made as well.

The RRF runs show the first significant improvements. The fusion run of the three runs BM25+Bo1, DFR $\chi^2$, and PL2 significantly outperform the BM25 baseline on MAP and nDCG to some extent. Larger improvements and the overall best results are achieved with BM25+monoT5. This run is significantly better on all measures and archives a 0.0324 higher nDCG. The passage retrieval version of the run performs considerably worse, similar to the baseline. 
The gap between the BM25 results on the two Doc2Query extended indexes is similar. While the results on the version with three additional queries per document make statistically no difference to the baseline, the results on the ten queries indexes are almost as good as the ones with BM25+monoT5 on all measures, except for P@20, which is even better. 
BM25+ColBERT performs slightly worse overall. Focusing on P@20, the system differs not from the baseline. Employing ColBERT as a first-stage ranker impairs the performance further. 
The results achieved with the E5 models as first-stage rankers are not significantly different from the baseline. Still, the base version outperforms the baseline in all measures, and the small version does on Bpref and RR.

\begin{figure}
  \centering
  \includegraphics[width=\linewidth]{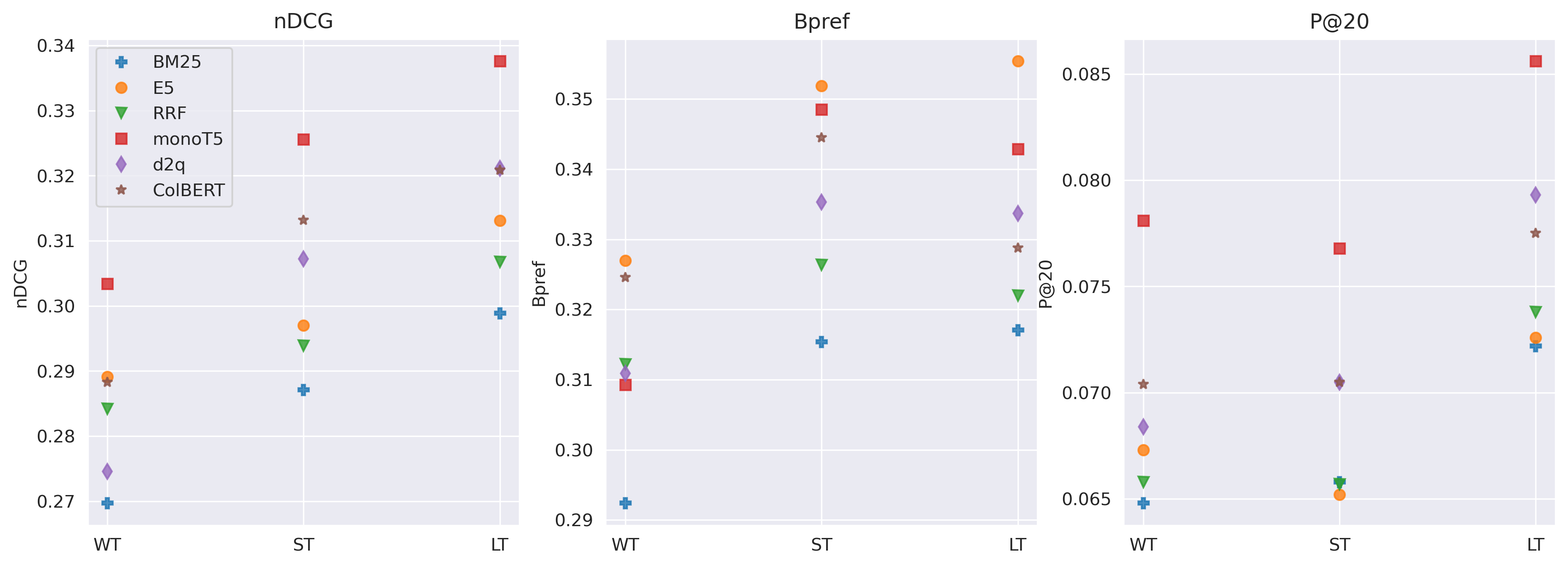}
  \caption{The ARP of nDCG (left), Bpref (center), and Reciprocal Rank (right) from the submitted systems at WT, ST, and LT.}
  \label{fig:results}
\end{figure}

\subsection{Results}
For the evaluation of the result, the main goal is not a high but rather persistent performance. The underlying assumption is that the system would continuously achieve the same performance. To evaluate this, the Result Delta ($\mathcal{R}_e\Delta$) between the averaged retrieval performances at two different points in time is measured as proposed by Sáez et al.~\cite{saezEvaluationInformationRetrieval2021}. The results are presented in Table~\ref{tab:results} and visualized in Figure~\ref{fig:results}.

\begin{table}[]
\caption{Results on the three (test) sub-collections as well as the deltas between them. The best system per measure and group is highlighted in \textbf{bold}, and significant differences from the BM25 baseline are denoted with an asterisk*.}
\label{tab:results}
\begin{tabular}{cl|lll|cc}
\toprule
        & & \multicolumn{3}{c}{ARP} & \multicolumn{2}{c}{$\mathcal{R}_e\Delta$} \\
        &         & WT      & ST      & LT      & \multicolumn{1}{l}{WT, ST} & \multicolumn{1}{l}{WT, LT} \\  \midrule

\multirow{6}{*}{\rotatebox[origin=c]{90}{Bpref}}        & BM25    & 0.2924  & 0.3154  & 0.3171  & -0.0230                   & -0.0247                   \\
                                                        & RRF     & 0.3122  & 0.3264* & 0.3220  & \textbf{-0.0142}                   & -0.0098                   \\
                                                        & ColBERT  & 0.3246  & 0.3445* & 0.3288  & -0.0392                   & -0.0336                   \\
                                                        & monoT5     & 0.3093  & 0.3485* & 0.3429* & -0.0244                   & -0.0228                   \\
                                                        & d2q & 0.3109  & 0.3353* & 0.3337* & -0.0199                   & \textbf{-0.0042}                   \\
                                                        & E5      & \textbf{0.3270}  & \textbf{0.3519*} & \textbf{0.3554*} & -0.0249                   & -0.0284                   \\ \midrule

\multirow{6}{*}{\rotatebox[origin=c]{90}{P@20}}         & BM25    & 0.0648  & 0.0658  & 0.0722  & -0.0010                   & -0.0074                   \\
                                                        & RRF     & 0.0658  & 0.0657  & 0.0738  & \textbf{0.0001}                    & -0.0080                   \\
                                                        & ColBERT  & 0.0704  & 0.0705* & 0.0775* & 0.0013                    & -0.0075                   \\
                                                        & monoT5     & \textbf{0.0781*} & \textbf{0.0768*} & \textbf{0.0856*} & -0.0021                   & -0.0109                   \\
                                                        & d2q & 0.0684  & 0.0705* & 0.0793* & \textbf{-0.0001}                   & -0.0071                   \\
                                                        & E5      & 0.0673  & 0.0652  & 0.0726  & 0.0021                    & \textbf{-0.0053}                   \\ \midrule
\multirow{6}{*}{\rotatebox[origin=c]{90}{nDCG}}         & BM25    & 0.2697  & 0.2871  & 0.2989  & -0.0174                   & -0.0292                   \\
                                                        & RRF     & 0.2842* & 0.2939* & 0.3068* & -0.0097                   & \textbf{-0.0226}                   \\
                                                        & ColBERT  & 0.2883  & 0.3132* & 0.3209* & -0.0222                   & -0.0342                   \\
                                                        & monoT5     & \textbf{0.3034}  & \textbf{0.3256*} & \textbf{0.3376*} & -0.0326                   & -0.0465                   \\
                                                        & d2q & 0.2746  & 0.3072* & 0.3211* & -0.0249                   & -0.0326                   \\
                                                        & E5      & 0.2891  & 0.2970  & 0.3131  & \textbf{-0.0079}                   & -0.0240                   \\
                                                        \bottomrule
\end{tabular}
\end{table}

\textbf{\texttt{IRC\_RRF(BM25+Bo1-XSqrA\_M-PL2)}:}
The fused run contains at least 1000 results for all topics in the WT sub-collection. For the ST sub-collection the system could not find any documents for four queries. Namely the queries \textit{to}, \textit{a}, \textit{the} and \textit{the}\footnote{LongEval ST qid: q072214697, q072222604, q072224942, q072212314} resulted in empty rankings. These queries consist only of stopwords, which leave an empty query string after query processing. These queries are most likely bad translations from the terms \textit{verseau, argentique, nanterre} and \textit{falloir}, mostly containing named entities.
For the two LT sub-collection topics \textit{cadreemploi} and \textit{a}\footnote{LongEval LT qid: q0922511 and q092219105}, no BM25 first stage ranking could be created. While \textit{a} is again just a stopword, for the term \textit{cadreemploi} no results were found, which could possibly be explained by a spelling error of the French job exchange website \textit{cadremploi}. Similarly, the topic \textit{cadreemploi} is also present in the French queries.

The Average Retrieval Performance (ARP)  --- defined by the mean retrieval performance over multiple topics ---  improves slightly over time. In general, the measured differences between the sub-collections are fairly small. The $\Delta$~nDCG between WT and ST is only -0.0097 and between WT and LT -0.0226. 

\textbf{\texttt{IRC\_BM25+colBERT}:}
Based on the WT sub-collection for the topic \textit{ducielalaterre}\footnote{LongEval WT held out qid: q062216851} no documents were found, and for all other topics, at least 1000 documents could be retrieved. 
Since ColBERT was employed as a reranker on top of BM25, the four topics \textit{to}, \textit{a}, \textit{the} and \textit{the}\footnote{LongEval ST qid: q072214697, q072222604, q072224942, and q072212314} still remain empty. For 28 other topics, only less than 1000 documents, ranging between three and 663, could be found.
Like before, the LT sub-collection topics \textit{cadreemploi} and the topic \textit{a}\footnote{LongEval LT qid: q0922511, q092219105} remain empty. For further 22 topics, less than 1000 results were found. For example, the fewest results were found for the topic \textit{the audeau}.\footnote{LongEval LT qid: q092220802}

The ARP is increasing over time, as already observed for the RRF system. However, the differences are larger for this system. Between WT and ST the $\Delta$~nDCG is -0.0249, and between WT and LT -0.0326. 

\textbf{\texttt{IRC\_BM25+monoT5}:}
The composition of the runs stayed mostly the same for these runs. Since they also use BM25 as the first-stage ranking, the issue of empty or short rankings remains. 

As already observed on the train slice of the WT sub-collection, the ARP is the highest achieved on all measures and sub-collections compared to the other submitted systems, with small exceptions. One strong exception is the Bpref of only 0.3093 on the WT sub-collection, the smallest score achieved overall. However, the results are inconsistent; the deltas are higher, especially for Bpref.

\textbf{\texttt{IRC\_d2q(10)>BM25}:}
Through the document expansion with Doc2Query, at least 37 documents were found for the previously empty WT sub-collection topic \textit{ducielalaterre}.\footnote{LongEval WT held out qid: q062216851} However, for the other sub-collections, the results stayed similar. 
Doc2Query performed weaker than initially on the train slice before, especially in comparison to monoT5. The result deltas between WT and ST and WT and LT are among the highest for nDCG and P@20. 

\textbf{\texttt{IRC\_E5\_base}:}
Since the E5 model is based on k-NN and no stopwords were removed, for every topic, 1000 results were found. 
Compared to the train slice of the WT sub-collection, the system performed better. It achieved the highest Bpref on all three sub-collections and a high overall nDCG. The results are especially consistent between sub-collections with a $\Delta$~nDCG of 0.0079 between WT and ST and -0.0240 between WT and LT.

\begin{figure}
  \centering
    \includegraphics[width=\linewidth]{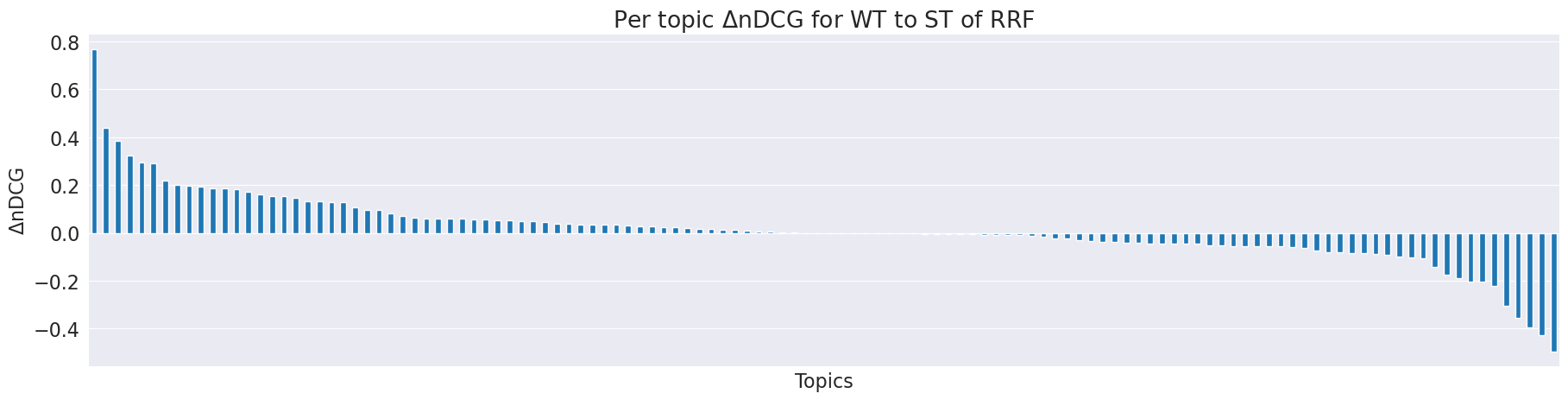}    
    \includegraphics[width=\linewidth]{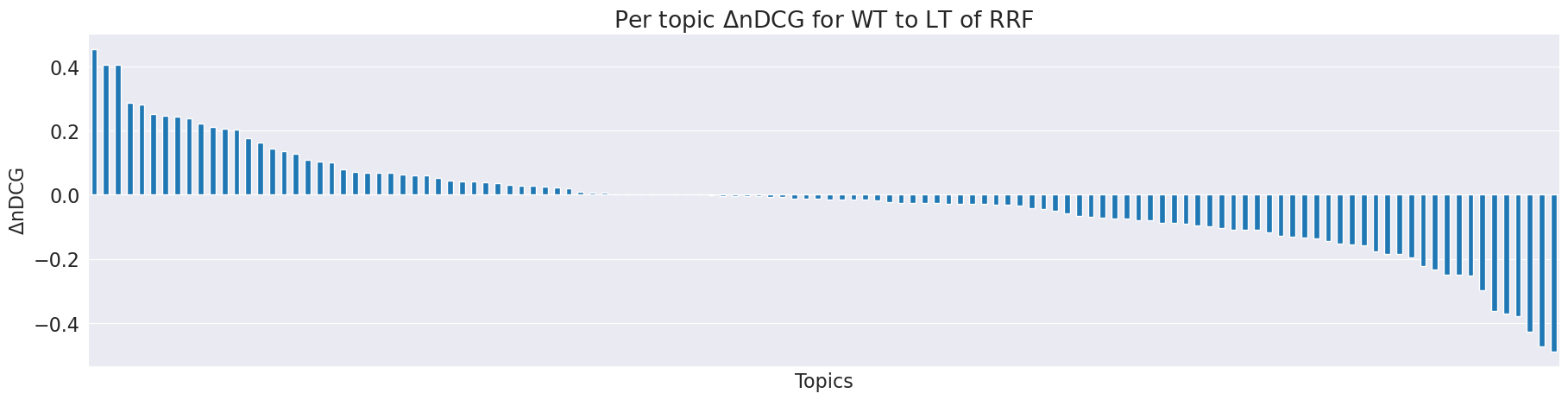}
  \caption{RRF $\Delta nDCG$ results per topic for WT to ST (top) and WT to LT (bottom). The topics are ordered according to the delta.}
  \label{fig:WTRRF}%
\end{figure}

\section{Temporal Persistence as Replicability}\label{sec:replicability}
Building upon the result delta evaluation as introduced by Sáez et al.~\cite{saezEvaluationInformationRetrieval2021}, we propose to use replicability measures to investigate the environment effect on the systems further. As described and implemented by Breuer et al.~\cite{breuerHowMeasureReproducibility2020a,breuerReproEvalPython2021a}, the ARP may hide differences between the topic score distributions. For example, the RRF system achieved a high nDCG (0.28) at WT and is relatively stable considering the  $\mathcal{R}_e\Delta(WT, ST)$ of 0.001. However, the per-topic results fluctuate between -0.4 and 0.8, as shown in Figure~\ref{fig:WTRRF}. For some topics, the retrieval performance improves, while the changes in the EE harm retrieval performance for other topics. We note that these circumstances require a more in-depth evaluation.

For a more detailed analysis of how the topic score distributions change, we cast the temporal comparison into a replication task, i.e., we evaluate the same set of systems on different data. Naturally, a direct comparison based on different sub-collections is difficult since it remains unclear if the observed effects should be attributed to the system or the changing EE. To overcome this problem, a pivot system similar to that described by Sáez et al.~\cite{saezEvaluationInformationRetrieval2021} is used, and likewise, the experimental system is kept fixed in both EE. Effects are measured in comparison to this pivot system on one sub-collection and then compared to the same setup on a later sub-collection. To align the terminology, the pivot system is a baseline run, BM25 for simplicity in this example, and the advanced run is the experimental system investigated. 

In addition to the $\mathcal{R}_e\Delta$, as reported earlier in Table~\ref{tab:results}, we report the Effect Ratio (ER) and the Delta Relative Improvement ($\Delta$~RI). The ER~\cite{breuerHowMeasureReproducibility2020a} is originally defined by the ratio between relative improvements of an advanced run over a baseline run. The relative improvements are based on the per-topic improvements, which are adapted for changing EEs as follows:

\begin{equation}
  \Delta M^{EE_1}_j = M^{EE_1}_j (S) - M^{EE_1}_j (P),  \Delta' M^{EE_2}_j = M^{EE_2}_j (S) - M^{EE_2}_j (P)
\end{equation}

where $\Delta M^{EE_1}_j$ denotes the difference in terms of a measure $M$ between the pivot system $P$ and the experimental system $S$ for the $j$-th topic of the evaluation environment $EE_1$. Correspondingly, $\Delta' M^{EE_2}_j$ denotes the topic-wise improvement in the evaluation environment $EE_2$. The ER is then defined as:

\begin{equation}
  \mathrm{ER} \big(\Delta'M^{EE_2}, \Delta M^{EE_1}\big) = \frac{\overline{\Delta'M^{EE_2}}}{\overline{\Delta M^{EE_1}}}= \frac{\frac{1}{n_{EE_2}}\sum^{n_{EE_2}}_{j=1}\Delta'M^{EE_2}_j}{\frac{1}{n_{EE_1}}\sum^{n_{EE_1}}_{j=1}\Delta M^{EE_1}_j}.
\end{equation}
\vspace{0.5em}

More specifically, the mean improvement per topic between the pivot and experimental system on one sub-collection (of $EE_1$) in comparison to the effect on the other sub-collection (of $EE_2$) is measured. Thereby, the ER is sensitive to the effect size. If the effect size is completely replicated in the second sub-collection, the ER is 1, i.e., the retrieval system is robust. If the ER is between 0 and 1, the effect is smaller, indicating a less robust system with performance drops. If the ER is larger than 1, the effect is larger, indicating performance gains caused by the change of the EE. Additionally, we include the $\Delta$~RI~\cite{breuerHowMeasureReproducibility2020a}, based on the relative improvements (RI) that are adapted to the LongEval definitions as follows:

\begin{equation}
    \mathrm{RI} = \frac{\overline{M^{EE_1}(S)-M^{EE_1}(P)}}{\overline{M^{EE_1}(P)}}, \qquad \mathrm{RI'}  = \frac{\overline{M^{EE_2}(S)-M^{EE_2}(P)}}{\overline{M^{EE_2}(P)}}
\end{equation}

where $M^{EE}$ denotes the score of a measure $M$ determined with $EE$, and $S$ and $P$ denote the experimental and pivot system, respectively. The $\Delta$~RI is then defined as:

\begin{equation}
    \Delta \mathrm{RI}= \mathrm{RI} - \mathrm{RI}'.
\end{equation}

Therefore, a comparison between different sub-collections is straightforward. The ideal $\Delta$~RI of 0 is achieved if the RI is the same between both sub-collections, indicating a robust system. The more $\Delta$~RI deviates from 0, the less robust is the system, whereas negative scores indicate a more effective experimental system $S$ in the evaluation environment $EE_2$, and higher scores correspond to a less effective experimental systems than in the evaluation environment $EE_1$. All of the replicability measures were implemented with the help of \texttt{repro\_eval}~\cite{breuerReproEvalPython2021a}, which is a dedicated reproducibility and replicability evaluation toolkit.

Even though the replicability measures do not necessarily require the same topics for each sub-collection, we harmonized the topics. Therefore, we only rely on the core queries that are shared between the sub-collections in this analysis.
Given this methodology, the extended results are presented in Table~\ref{tab:replicability}. For all systems, the ARP decreases slightly at first (WT to ST) but increases in the long run (WT to LT) --- a circumstance that is also reflected by the lower $\mathcal{R}_e\Delta$ scores for WT to ST compared to WT to LT. 

The ER and $\Delta$~RI complement $\mathcal{R}_e\Delta$. For instance, monoT5 achieved similar P@20 scores on WT and ST, resulting in a $\mathcal{R}_e\Delta$ score of 0, which indicates perfect robustness in terms of $\mathcal{R}_e\Delta$. However, when comparing ER and also $\Delta$~RI, more granular analysis is possible. In this case, the scores are close to but different from the perfect scores of 1 and 0, respectively, which would indicate perfect robustness. In general, the $\mathcal{R}_e\Delta$ scores do not always agree on the most robust system with ER and $\Delta$~RI. By these findings, we conclude that the replicability measures provide another perspective of the robustness, and we emphasize once again that it is also important to consider the topical variance over time.

Furthermore, we see that it is not enough to consider the differences of a single retrieval measure like nDCG. Depending on the evaluation measure, different systems perform best in terms of robustness. For instance, $\mathcal{R}_e\Delta$ of nDCG is lower for ColBERT and Doc2Query than that of monoT5, while $\mathcal{R}_e\Delta$ of P@20 is lower for monoT5. Similarly, the replicability measures should be instantiated with different retrieval measures to get a more comprehensive understanding of robustness. While our RRF-based submissions achieve the best $\mathrm{ER}_{\mathrm{nDCG}}$ on both tasks, monoT5 is the most robust system in terms of $\mathrm{ER}_{\mathrm{P@20}}$. Likewise, ER and $\Delta$~RI identify different systems as the most robust for the same measures and tasks, which shows that it is insightful to evaluate both replicability measures.

In addition, we also included the p-values of unpaired tests based on the topic score distributions from different EE that were determined with the same experimental system as proposed in~\cite{breuerHowMeasureReproducibility2020a}. The general idea of these evaluations proposes to determine the quality of replicability (in our case, robustness) by the p-values. It follows the assumption that lower p-values give a higher probability of failed replications or systems that are not robust. As can be seen, the highest p-values are achieved for the monoT5, ColBERT, or d2q, which generally agrees with our earlier observations.

\begin{table}[]
        \caption{Extended results on the core queries, including the replicability measures.}
        \label{tab:replicability}
        \resizebox{\textwidth}{!}{
        \begin{tabular}{cl|rrr|rr|rr|rr|rr}
        \toprule
        \multicolumn{2}{l}{}                                                        &\multicolumn{3}{c}{ARP}        & \multicolumn{2}{c}{$\mathcal{R}_e\Delta$}   & \multicolumn{2}{c}{ER} & \multicolumn{2}{c}{$\Delta$~RI} & \multicolumn{2}{c}{p-val} \\
                                                        & System  & WT              & ST                & LT                & WT, ST            & WT, LT             & WT, ST   & WT, LT & WT, ST   & WT, LT & WT, ST & WT, LT   \\ \midrule
        \multirow{6}{*}{\rotatebox[origin=c]{90}{P@20}} & BM25    & 0.070           & 0.067             & 0.085             & 0.002             & -0.015             &  1.000       & 1.000      &  0.000       &  0.000     &   1.000       & 1.000       \\ 
                                                        & RRF     & 0.075           & 0.069             & 0.088             & 0.006             & \textbf{-0.013}    & 0.311    & 0.544  & 0.051    & 0.041  & 0.591    & 0.269  \\
                                                        & colBERT & 0.072           & 0.071             & 0.087             & 0.002             & -0.015             & 1.244    & 0.933  & \textbf{-0.011}   & \textbf{0.009}  & 0.875    & 0.190  \\
                                                        & monoT5  & \textbf{0.081}  & \textbf{0.081}    & \textbf{0.096}    & \textbf{0.000}    & -0.014             &\textbf{ 1.191}    & \textbf{0.953}  & -0.039   & 0.037  & \textbf{0.998}    & 0.229  \\
                                                        & d2q     & 0.079           & 0.072             & 0.091             & 0.007             & \textbf{-0.013}    & 0.499    & 0.726  & 0.062    & 0.051  & 0.547    & \textbf{0.303}  \\
                                                        & E5      & 0.071           & 0.066             & 0.088             & 0.005             & -0.017             & -1.452   & 2.903  & 0.040    & -0.022 & 0.616    & 0.125  \\ \midrule
        \multirow{6}{*}{\rotatebox[origin=c]{90}{nDCG}} & BM25    & 0.269           & 0.272             & 0.306             & -0.003            & -0.037             & 1.000        &   1.000    &   0.000      & 0.000      &   1.000      & 1.000      \\
                                                        & RRF     & 0.285           & 0.282             & 0.314             & 0.003             & -0.030             & \textbf{0.925}    & \textbf{0.786}  & \textbf{0.003}    & 0.013  & 0.945    & 0.227  \\
                                                        & colBERT & 0.276           & 0.275             & 0.297             & \textbf{0.001}    & -0.021             & 0.441    & -1.198 & 0.015    & 0.053  & \textbf{0.967}    & 0.412  \\
                                                        & monoT5  & \textbf{0.295}  & \textbf{0.302}    & 0.311             & -0.007            & \textbf{-0.015}    & 1.146    & 0.187  & -0.013   & 0.083  & 0.817    & \textbf{0.580}  \\
                                                        & d2q     & 0.285           & 0.287             & \textbf{0.327}    & \textbf{-0.001}   & -0.042             & 0.916    & 1.317  & 0.006    & \textbf{-0.010} & 0.960    & 0.150  \\
                                                        & E5      & 0.290           & 0.300             & 0.313             & -0.010            & -0.023             & 1.333    & 0.362  & -0.025   & 0.054  & 0.720    & 0.382  \\ \midrule
        \multirow{6}{*}{\rotatebox[origin=c]{90}{Bpref}}& BM25    & 0.314           & 0.314             & 0.324             & \textbf{-0.000}   & -0.010             &  1.000       &    1.000   &  0.000       &  0.000     &  1.000       &  1.000     \\ 
                                                        & RRF     & 0.346           & 0.328             & 0.347             & 0.019             & -0.001             & 0.574    & \textbf{1.007}  & 0.032    & \textbf{0.002}  & 0.784    & 0.756  \\
                                                        & colBERT & 0.324           & 0.317             & 0.338             & 0.007             & -0.013             & 0.286    & 1.278  & 0.024    & -0.008 & 0.826    & 0.668  \\
                                                        & monoT5  & 0.337           & 0.344             & 0.337             & -0.007            & \textbf{0.000}     & 1.261    & 0.553  & -0.019   & 0.034  & 0.850    & \textbf{0.997}  \\
                                                        & d2q     & 0.335           & 0.331             & 0.368             & 0.004             & -0.033             & \textbf{0.779}    & 2.034  & \textbf{0.015}    & -0.067 & \textbf{0.894}    & 0.300  \\
                                                        & E5      & \textbf{0.368}  & \textbf{0.354}    & \textbf{0.371}    & 0.014             & -0.003             & 0.738    & 0.863  & 0.045    & 0.028  & 0.692    & 0.931  \\ \bottomrule
        \end{tabular}
        }
        \end{table}

\begin{figure}%
    \centering
    \includegraphics[width=0.43\linewidth]{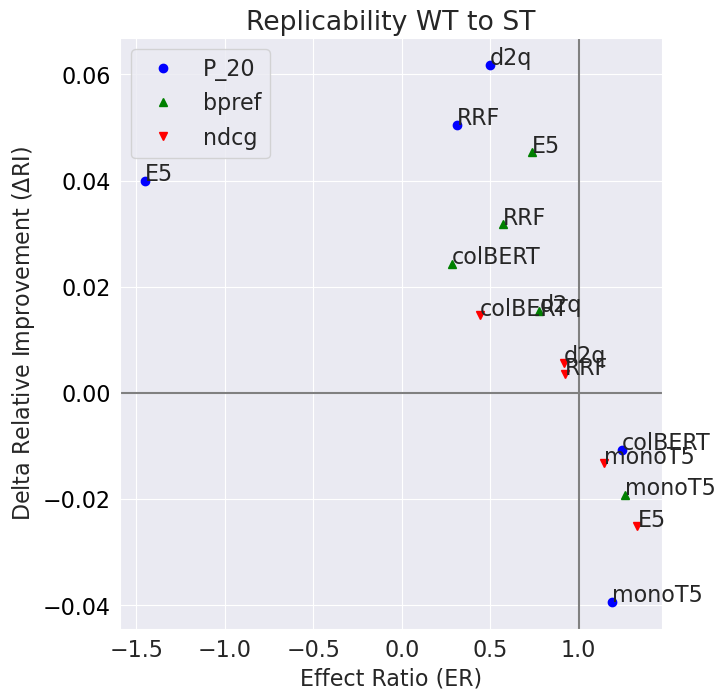}
    \qquad
    \includegraphics[width=0.4\linewidth]{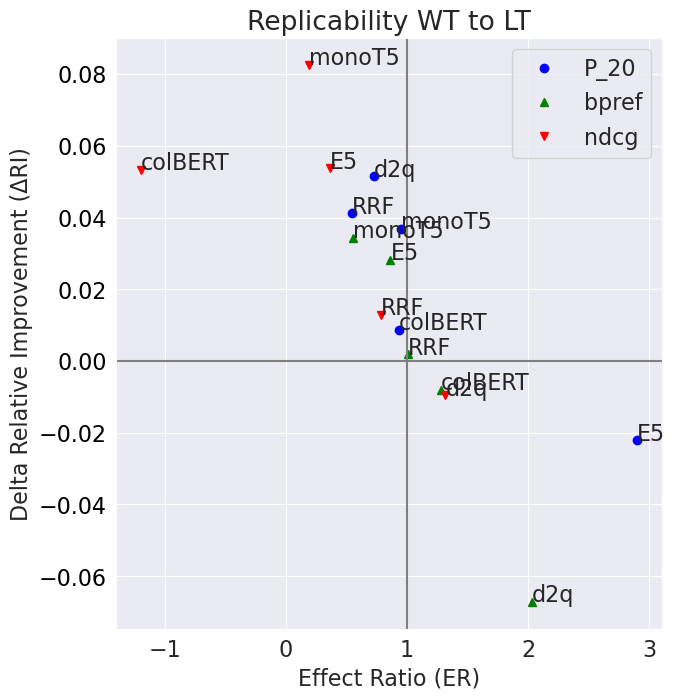}
    \caption{The ER plotted against the $\Delta$~RI for the replication WT to ST (left) and WT to LT (right).}%
    \label{fig:DRI-ER}%
\end{figure}

The full potential of the ER and $\Delta$~RI can be seen if plotted against each other as in Figure~\ref{fig:DRI-ER}. The closer the systems are located to the point (1, 0), the more persistent they are, with the preferable regions bottom right and top left. For the comparison of WT to ST, the monoT5 system performs well on all three measures. However, the effect and the absolute scores are larger. The E5 system completely fails to replicate the absolute P@20 score and shows a generally larger difference. The RRF system, like most others, shows smaller absolute scores according to the $\Delta$~RI and a slightly decreased effect ratio. The plot regarding WT to LT shows more outliers with larger effect sizes for P@20 for the E5 system and Bpref for the d2q system. The systems are shifted to the top right of the plot, a trend similar to the increased $\mathcal{R}_e\Delta$ for WT to LT. 

\section{Conclusion and Outlook}\label{sec:conclusion}
In this work, we described our participation in the LongEval Lab at CLEF 2023. As the core contribution, we applied five advanced retrieval systems to the LongEval dataset and submitted the runs to both sub-tasks. As this is a new challenge, the interpretation of the results is difficult. The results for the different systems are very similar. The measured differences are statistically significant but appear small as compared to the same methods on different datasets as listed on the IR experiment platform~\cite{froebe:2023e}.\footnote{\url{https://www.tira.io/task/ir-benchmarks}} Interestingly, an increasing ARP over time was observed for most systems and measures. Still, the performance difference, measured by $\mathcal{R}_e\Delta$, is smaller for WT to ST compared to WT to LT, which complies with the natural assumption that persistence deteriorates over time.

Further, we report preliminary results applying replicability measures to quantify temporal persistence, an extension on common practices of these measures and their interpretation~\cite{DBLP:journals/ipm/MaistroBSF23}. It was shown that the results based on different measures and likewise for different topics do not necessarily agree with each other. Therefore, we see great potential in using replicability measures to gain further insights into robustness and also saw similarities to the measured result deltas. All in all, a strong environment effect on the systems was shown and could be analyzed.

Future work will be regarding the selection of the pivot system and qualitative core queries. Also, further harmonizing the dataset by unifying the document IDs would allow us to cast the problem as a reproducibility task and investigate persistence on an even more specific level with reproducibility measures.

\bibliography{ceur}



\end{document}